\documentclass[aps,preprint,nofootinbib,eqsecnum]{revtex4}
\usepackage{amsmath,amssymb,bm}
\usepackage{graphicx}
\usepackage{epstopdf}
\usepackage{latexsym}

\begin{document}
\preprint{arXiv:0806.0002}

\title{Renormalization group theory of nematic ordering\\ in $d$-wave superconductors}
\author{Yejin Huh}
\affiliation{Department of Physics, Harvard University, Cambridge MA 02138, USA}
\author{Subir Sachdev}
\affiliation{Department of Physics, Harvard University, Cambridge MA 02138, USA}
\date{May 31, 2008\\
\vspace{1.6in}}

\begin{abstract}
~\\
We examine the quantum theory of the spontaneous breaking of lattice rotation symmetry
in $d$-wave superconductors on the square lattice. This is described by a field theory of an
Ising nematic order parameter coupled to the gapless fermionic quasiparticles. We determine
the structure of the renormalization group to all orders in a $1/N_f$ expansion, where $N_f$ is
the number of fermion spin components. Asymptotically exact results are obtained for
the quantum critical theory in which, as in the large $N_f$ theory, 
the nematic order has a large anomalous dimension, 
and the fermion spectral functions are highly anisotropic.
\end{abstract}

\maketitle

\section{Introduction}

There has been a great deal of research on the onset of a variety of competing orders
in the hole-doped cuprate superconductors. In Ref.~\onlinecite{vojta}, a classification of
spin-singlet order parameters at zero momentum was presented: such orders are able to couple
efficiently to the gapless nodal quasiparticle excitations of a $d$-wave superconductor. 
Our focus in the present paper will be on one such order parameter: `nematic' ordering in
which the square lattice (tetragonal) symmetry of the $d$-wave superconductor is spontaneously
reduced to rectangular (orthorhombic) symmetry \cite{fke}. This transition is characterized by an
Ising order parameter, but the quantum phase transition is not in the usual Ising universality class \cite{vojta}
because
the coupling to the gapless fermionic quasiparticles changes the nature of the quantum critical fluctuations.
Our work is motivated by recent neutron scattering observations \cite{hinkov} of a strongly temperature ($T$)
dependent susceptibility to nematic ordering in detwinned crystals of YBa$_2$Cu$_3$O$_{6.45}$.

An initial renormalization group (RG) analysis \cite{vojta} of nematic ordering at $T=0$ found runaway flow to strong-coupling
in a computation based on an expansion in $(3-d)$ where $d$ is the spatial dimensionality.
More recently, Ref.~\onlinecite{eakim} argued that a second-order quantum phase transition existed in the limit
of large $N_f$, where $N_f$ is the number of spin components (the physical case corresponding to 
$N_f = 2$). We refer the reader to Ref.~\onlinecite{eakim} for further discussion on the physical importance
and experimental relevance of nematic ordering in $d$-wave superconductors. 

\begin{figure}
\includegraphics[width=2.3in]{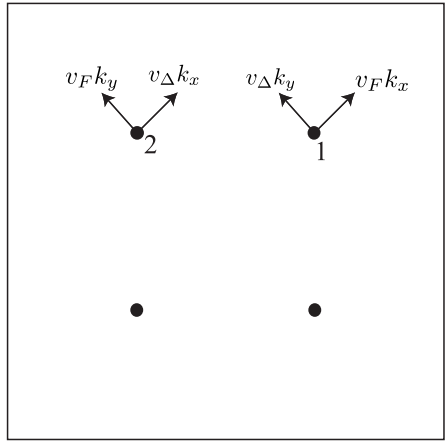}
\caption{Nodal points of the $d$-wave superconductor in the square lattice Brillouin zone.
The 2-component $\Psi_{1,2}$ fermions are in the vicinity of the labelled nodal points and their partners
at diagonally opposite points.} \label{fig:nodal}
\end{figure}

Here we will present a RG analysis using the framework of the $1/N_f$ expansion. 
We will find that a fixed point does indeed exist at order $1/N_f$, describing a second-order quantum transition
associated with the onset of long-range nematic order.
However, the scaling properties near the fixed point have a number of unusual properties,
controlled by the marginal flow of a ``dangerously'' irrelevant parameter. This parameter is $v_\Delta/v_F$,
where $v_{\Delta}$ and $v_F$ are the velocities of the nodal fermions parallel and perpendicular to the Fermi 
surface (see Fig.~\ref{fig:nodal}). 
We will show that the fixed point has $(v_\Delta /v_F)^\ast = 0$ and so the
transition is described by an ``infinite anisotropy'' away from the ``relativistic'' fixed points found for other
competing orders \cite{vojta}. It is important to note, however, that even though the fixed point has $(v_\Delta /v_F)^\ast = 0$,
it is not described by an effectively one-dimensional theory of a straight Fermi surface; a fully two-dimensional
theory is needed as is discussed in more detail in Section~\ref{sec:ho}.
 The approach to this fixed point is logarithmically slow, and physical properties have
to be computed at finite $v_\Delta /v_F$. Our main results for the RG flow of the velocities are in Eqs.~(\ref{vfl}) and
(\ref{vdl}), where $C_{1,2,3}$ are functions only of $v_\Delta/v_F$ which are specified in Eq.~(\ref{rg5});
for $v_\Delta/v_F \ll 1$, the equations reduce to the explicit forms in Eqs.~(\ref{rgf1}-\ref{rgf3}).
These (and related) equations can be integrated in a standard manner to yield the dependence of observables
on temperature and deviation from the nematic critical point, and the results are in Figs.~\ref{fig:vplot1}, \ref{fig:vplot2},
and \ref{fig:vplot3}.

The existence of a fixed point at $(v_\Delta /v_F)^\ast = 0$ suggests that we analyze the theory directly 
in the limit of small $v_\Delta / v_F$, without an appeal to an expansion in powers of $1/N_f$. The nature of the small
$v_\Delta / v_F$ limit is quite subtle, and has to be taken with care: it will be described in Section~\ref{sec:ho}.
We argue there that the fluctuations of the nematic order are controlled by a small parameter which is $\sim v_\Delta /(N_f v_F)$,
and not $1/N_f$ alone. Consequently, we believe our computations are controlled in the limit of small $v_\Delta /v_F$ even for
$N_f = 2$. Indeed, the results in Eqs.~(\ref{rgf1}-\ref{rgf3}), and many other related results, are expected to be exact as we
approach the quantum critical point.

The outline of the remainder of this paper is as follows. The field theory for the nematic ordering transition
will be reviewed in Section~\ref{sec:ft}, along with a discussion of the $1/N_f$ expansion. The RG analysis to order $1/N_f$ will
be presented in Section~\ref{sec:rg}. Finally, higher order corrections in $1/N_f$, and the nature of the small $v_\Delta/v_F$ limit
will be discussed in Section~\ref{sec:ho}

\section{Field theory}
\label{sec:ft}

We begin by review the field theory for the nematic ordering transition introduced
in Ref.~\onlinecite{vojta}.

The action for the field theory, $S$, has three components
\begin{equation}
S = S_\Psi + S_{\phi}^0 + S_{\Psi \phi}.
\end{equation}

The first term in the action, $S_{\Psi}$, is simply
that for the low energy fermionic excitations in the $d_{x^2-y^2}$
superconductor.
We begin with the electron annihilation operator $c_{{\bf q}a}$ 
at momentum ${\bf q}$ and spin $a = \uparrow, \downarrow$. We will shortly
generalize the theory to one in which $a= 1 \ldots N_f$, with $N_f$ an arbitrary integer.
We denote the $c_{{\bf q}a}$ operators 
in the vicinity of the four nodal
points ${\bf q} = (\pm K, \pm K)$
by (anti-clockwise) $f_{1a}$, $f_{2a}$, $f_{3a}$, $f_{4a}$.
Now introduce the 2-component Nambu spinors $\Psi_{1a} =
(f_{1a}, \varepsilon_{ab} f_{3b}^{\dagger})$
and  $\Psi_{2a} =
(f_{2a}, \varepsilon_{ab} f_{4b}^{\dagger})$ where
$\varepsilon_{ab}=-\varepsilon_{ba}$ and
$\varepsilon_{\uparrow \downarrow} = 1$ [we will follow the
convention of writing out spin indices ($a,b$) explicitly, while indices
in Nambu space will be implicit].
Expanding to linear order in gradients from the nodal points,
the Bogoliubov action for the fermionic excitations of a $d$-wave superconductors
can be written as (see Fig.~\ref{fig:nodal})
\begin{eqnarray}
S_{\Psi} &=& \int \!\! \frac{d^2 k}{(2 \pi)^2} T \!\sum_{\omega_n} \sum_{a = 1}^{N_f}
\Psi_{1a}^{\dagger}  \left(
- i \omega_n + v_F k_x \tau^z + v_{\Delta} k_y \tau^x \right) \Psi_{1a}  \nonumber \\
&+& \int \!\! \frac{d^2 k}{(2 \pi)^2}
T \! \sum_{\omega_n}  \sum_{a = 1}^{N_f}
\Psi_{2a}^{\dagger}  \left(
- i \omega_n + v_F k_y \tau^z + v_{\Delta} k_x \tau^x \right) \Psi_{2a} .
\label{dsid1}
\end{eqnarray}
Here $\omega_n$ is a Matsubara frequency,
$\tau^{\alpha}$ are Pauli matrices which act in the fermionic
particle-hole space, $k_{x,y}$ measure the wavevector from the nodal points and
have been rotated
by 45 degrees from $q_{x,y}$ co-ordinates,
and $v_{F}$, $v_{\Delta}$ are velocities. The sum over $a$ in Eq.~(\ref{dsid1})
can be considered to extend over an arbitrary number $N_f$.

The second term, $S_{\phi}^0$ describes the effective action for the
Ising nematic order parameter, which we represent as a real field $\phi$. Considering only
the contribution to its action generated by high energy electronic excitations, we obtain
only the analytic terms present in the quantum Ising model:
\begin{equation}
S_{\phi}^0 = \int \!\! d^2 x d \tau \Big[
\frac{1}{2}(\partial_{\tau} \phi)^2 + \frac{c^2}{2} (\nabla \phi )^2 +
\frac{r}{2} \phi^2 + \frac{u_0}{24} \phi^4 \Big];
\label{dsid3}
\end{equation}
here $\tau$ is imaginary time,
$c$ is a velocity, $r$ tunes the system across the
quantum critical point, and $u_0$ is a quartic self-interaction.

The final term in the action, $S_{\Psi\phi}$ couples the Ising nematic order, $\phi$, to 
the nodal fermions, $\Psi_{1a}$, $\Psi_{2a}$. This can be deduced by a symmetry analysis \cite{vojta},
and the important term is a trilinear ``Yukawa'' coupling
\begin{equation}
S_{\Psi\phi} = \int \!\! d^2 x d \tau \Big[ \lambda_0 \phi
\sum_{a = 1}^{N_f}\left( \Psi_{1a}^{\dagger} \tau^x \Psi_{1a} + \Psi_{2a}^{\dagger} \tau^x
\Psi_{2a} \right) \Big],
\label{dsid4}
\end{equation}
where $\lambda_0$ is a coupling constant.

We can now see that the action $S$ describes the couplings between the $\phi$ bosons
and the $\Psi$ fermions, both of which have a `relativistic' dispersion spectrum. However, the velocity of `light' in 
their dispersion spectrums, $c$, $v_F$, $v_\Delta$, are not all equal. If we choose equal 
velocities with $v_F = v_\Delta = c$, then the decoupled theory $S_\Psi + S_\phi^0$ does have a relativistically
invariant form. However, even in this case, the fermion-boson coupling $S_{\Psi\phi}$ is {\em not\/}
relativistically invariant: the coupling matrix $\tau^x$ in Eq.~(\ref{dsid4}) breaks Lorentz symmetry,
and this feature will be crucial for our analysis. If we replace $\tau^x$ by $\tau^y$ in Eq.~(\ref{dsid4}),
then we obtain an interacting relativistically invariant theory for the equal velocity case, 
which was studied in Ref.~\onlinecite{vojta}
as a description of the transition from a $d_{x^2 - y^2}$ to a 
$d_{x^2 - y^2} + i d_{xy}$ superconductor. 

A renormalization group analysis of the above action has been presented previously \cite{vojta} in an
expansion in $(3-d)$. No fixed point describing the onset of nematic order was found, with
the couplings $u_0$ and $\lambda_0$ flowing away to infinity. This runaway flow was closely
connected to the non-Lorentz-invariant structure of $S_{\Psi\phi}$ and the resulting flow of the velocities
away from each other. In contrast, for the transition to a $d_{x^2 - y^2} + i d_{xy}$ superconductor, 
a stable fixed point was found at which the velocities flowed to equal values at long scales \cite{vojta}.
Similar relativistically-invariant fixed points are also found in other cases involving the onset
of spin or charge density wave orders at wavevectors which nest the separation between nodal points \cite{bfn,ying}.

\subsection{Large $N_f$ expansion}
\label{sec:nf}

We will analyze $S$ in the context of the $1/N_f$ expansion. Formally, this involves integrating out the $\Psi_{1,2}$ fermions,
and obtaining the resulting non-local effective action for $\phi$.
Near the potential quantum critical point, the non-local terms so generated are more important in the
infrared than the terms $S_\phi^0$ in Eq.~(\ref{dsid3}), apart from the ``mass'' term $r$. So we can drop the remaining
terms in $S_\phi^0$; also for convenience we rescale $\phi \rightarrow \phi/\lambda_0$ and $r \rightarrow N_f r \lambda_0^2 $, and 
obtain the local field theory
\begin{equation}
S = S_\Psi + \int d^2 x d \tau \left[ \frac{N_f r}{2} \phi^2 +  \phi
\sum_{a = 1}^{N_f} \left( \Psi_{1a}^{\dagger} \tau^x \Psi_{1a} + \Psi_{2a}^{\dagger} \tau^x
\Psi_{2a} \right) \right].
\label{local}
\end{equation}
This local field theory will form the basis of all our RG analysis. 
Note that this field theory has only 3 parameters: $r$, $v_F$, and $v_\Delta$. So any RG equations can be expressed
only in terms of these parameters, as will be presented in the following section.

\begin{figure}
\includegraphics[width=3in]{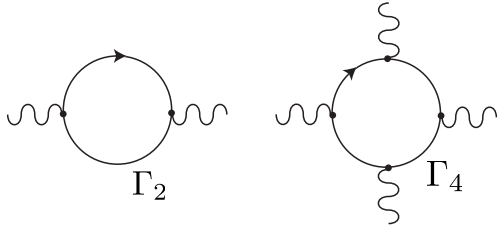}
\caption{Feynman graph expansion for the effective action $S_\phi$. The full lines are fermion propagators, while the wavy
lines are $\phi$ insertions.} \label{fig:feyn}
\end{figure}

The large $N_f$ expansion proceeds by integrating out the $\Psi$ fermions, yielding an effective action $S_\phi$ for the
nematic order $\phi$.
It is important to note that $S_\phi$ is a complicated non-local functional of $\phi$, which however
depends upon only $r$, $v_F$, and $v_\Delta$.
Also, in writing out explicit forms for $S_\phi$ it is essential to keep subtle issues on orders of limits
in mind. In particular, for the large $N_f$ phase diagram, we need the effective potential
for $\phi$ at zero $k$ and $\omega$. Thus we need an expansion of the effective potential in the regime
$|\phi| \gg |k|, |\omega|$ --- the structure of this was discussed in Ref.~\onlinecite{eakim}, and  
yielded a second-order transition in the limit of large $N_f$.
In contrast, for our RG analysis here, we will work in the quantum critical region, where $\langle \phi \rangle  =0$,
and so we need the functional for $|\phi| \ll |k|, |\omega|$. In this case, we can expand $S_\phi$ in powers of $\phi$
to yield the formal result
\begin{eqnarray}
\frac{S_\phi}{N_f} &=& \frac{1}{2} \int_K \left( r + \Gamma_2 (K) \right) |\phi (K)|^2 
\nonumber \\ &~&~~~~~+ \frac{1}{4} \prod_{i=1}^4 \int_{K_i} \delta\left(
\sum_i K_i \right)
\Gamma_4 (K_1,K_2,K_3,K_4) \phi (K_1) \phi (K_2) \phi (K_3) \phi (K_4) + \ldots
\label{sp1}
\end{eqnarray}
Here the $K_i \equiv (k_i, \omega_i)$ are 3-momenta.
The functions $\Gamma_i$ are all given by one fermion loop Feynman graphs with $i$
insertions of the external $\phi$ vertices, as shown in Fig.~\ref{fig:feyn}; we denote the external momenta of these vertices, $K_i$,
clockwise around the fermion loop. 
The Feynman loop integrals are quite tedious to evaluate, especially
for large $i$, and so below we only present explicit expressions for the needed low order
terms. However, all the $\Gamma_i$ are universal
functions of only the momenta and $v_F$ and $v_\Delta$

Our analysis will require the explicit form of $\Gamma_2 (K )$. This can be written as
\begin{equation}
\Gamma_2 (K) = \Pi_2 (k_x, k_y, \omega) + \Pi_2 (k_y, k_x, \omega)
\end{equation}
with the 2 terms representing the contributions of the $\Psi_1$ and $\Psi_2$ fermions
respectively. The one fermion loop diagram yields
\begin{eqnarray}
\Pi_2 (k_x , k_y, \omega) &=& \int \frac{d^2 p}{4 \pi^2} \int \frac{d \Omega}{2 \pi} \mbox{Tr} 
\Bigl[ \tau^x \left( - i (\Omega+ \omega) + v_F (p_x + k_x) \tau^z + v_\Delta ( p_y + k_y) \tau^x \right)^{-1} 
\nonumber \\
&~&~~~~~~\times \tau^x \left( - i \Omega + v_F p_x \tau^z + v_\Delta  p_y \tau^x \right)^{-1}
\Bigr] \nonumber \\
&=& \frac{1}{16 v_F v_\Delta}
 \frac{(\omega^2 + v_F^2 k_x^2)}{(\omega^2 + v_F^2 k_x^2 + v_\Delta^2 k_y^2)^{1/2}} \label{pi2r}
\end{eqnarray}
Here, we have subtracted out a constant which shifts the position of the critical point.

For $\Gamma_4$ we will only need the cases where 2 of the four external 3-momenta vanish. The first is
\begin{equation}
\Gamma_4 ( K,-K,0,0) = \Pi_{4a} (k_x, k_y, \omega) + \Pi_{4a} (k_y, k_x, \omega)
\end{equation}
where
\begin{eqnarray}
\Pi_{4a} (k_x , k_y, \omega) &=& \int \frac{d^2 p}{4 \pi^2} \int \frac{d \Omega}{2 \pi} \mbox{Tr} 
\Bigl[ \Bigl\{ \tau^x \left( - i (\Omega+ \omega) + v_F (p_x + k_x) \tau^z + v_\Delta ( p_y + k_y) \tau^x \right)^{-1}  \Bigr\}^3
\nonumber \\
&~&~~~~~~\times \tau^x \left( - i \Omega + v_F p_x \tau^z + v_\Delta  p_y \tau^x \right)^{-1}
\Bigr] \nonumber \\
&=& \frac{1}{2v_\Delta^2} \frac{\partial^2 \Pi_2 (k_x, k_y, \omega)}{\partial k_y^2} \nonumber \\
&=& -\frac{1}{32 v_F v_\Delta}
 \frac{(\omega^2 + v_F^2 k_x^2)(\omega^2 + v_F^2 k_x^2 -2 v_\Delta^2 k_y^2)}{(\omega^2 + v_F^2 k_x^2 + v_\Delta^2 k_y^2)^{5/2}}.
\label{p4a}
\end{eqnarray}
The other case with 2 vanishing 3-momenta is
\begin{equation}
\Gamma_4 ( K,0,-K,0) = \Pi_{4b} (k_x, k_y, \omega) + \Pi_{4b} (k_y, k_x, \omega)
\end{equation}
where
\begin{eqnarray}
\Pi_{4b} (k_x , k_y, \omega) &=& \int \frac{d^2 p}{4 \pi^2} \int \frac{d \Omega}{2 \pi} \mbox{Tr} 
\Bigl[ \Bigl\{ \tau^x \left( - i (\Omega+ \omega) + v_F (p_x + k_x) \tau^z + v_\Delta ( p_y + k_y) \tau^x \right)^{-1}  \Bigr\}^2
\nonumber \\
&~&~~~~~~\times \Bigl\{\tau^x \left( - i \Omega + v_F p_x \tau^z + v_\Delta  p_y \tau^x \right)^{-1} \Bigr\}^2
\Bigr] \nonumber \\
&=&  \frac{1}{16 v_F v_\Delta}
 \frac{(\omega^2 + v_F^2 k_x^2)(\omega^2 + v_F^2 k_x^2 -2 v_\Delta^2 k_y^2)}{(\omega^2 + v_F^2 k_x^2 + v_\Delta^2 k_y^2)^{5/2}}.
 \label{p4b}
\end{eqnarray}

From the effective action $S_\phi$ in (\ref{sp1}), we can obtain the $1/N_f$ corrections to various observable quantities.

For the nematic susceptibility, $\chi_\phi$, we have
\begin{equation}
\chi_\phi^{-1} = r + \frac{1}{N_f} \int_K \frac{2\Gamma_4 (K,-K,0,0) + \Gamma_4 (K,0,-K,0)}{\Gamma_2 (K) + r}.
\label{sp2}
\end{equation}
Using the values in Eq.~(\ref{p4a}) and (\ref{p4b}), we observe that the $1/N_f$ correction is identically zero.
This can be traced to the arguments in Ref.~\onlinecite{eakim} (based upon gauge invariance and 
the fact that the coupling in Eq.~(\ref{dsid4}) is to a globally conserved fermion current) that the effective potential
for the field $\phi$ is unrenormalized by the low energy fermion action. Indeed, this argument implies that 
all higher order terms in $1/N_f$ also vanish, and that $\chi_\phi^{-1} = r$ exactly in the present continuum field theory.
So we may conclude that the susceptibility exponent, $\gamma$, for the nematic order parameter has the exact value
\begin{equation}
\gamma = 1. \label{gamma}
\end{equation}
Note that there are $1/N_f$ corrections to the momentum and frequency dependence of the nematic susceptibility,
and these will appear in our RG analysis below.

\begin{figure}
\includegraphics[width=3in]{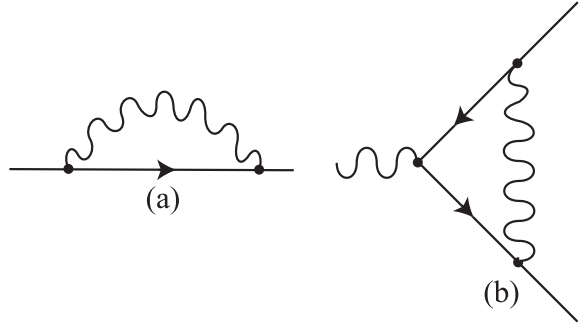}
\caption{Order $1/N_f$ contributions to the ({\em a\/}) self energy $\Sigma_1$ of $\Psi_1$ fermions
and ({\em b\/}) the vertex $\Xi$.} \label{fig:feyn2}
\end{figure}
Turning to the $1/N_f$ expansion for the $\Psi_1$ Green's function, we write this as
\begin{equation}
G_{\Psi_1}^{-1} (k, \omega) = -i \omega + v_F k_x \tau^z + v_{\Delta} k_y \tau^x - \Sigma_1 (k, \omega),
\end{equation}
the self-energy is given by the Feynman graph in Fig.~\ref{fig:feyn2}:
\begin{eqnarray}
\Sigma_1 (k, \omega) &=& \frac{1}{N_f} \int \frac{d^2 p}{4 \pi^2} \int \frac{d \Omega}{2 \pi} 
\frac{\left[  i(\Omega + \omega) - v_F (p_x + k_x) \tau^z + v_\Delta (p_y + k_y) \tau^x \right]}{(\Omega+ \omega)^2 + v_F^2 (p_x + k_x)^2 + v_\Delta^2 (p_y + k_y)^2} \nonumber \\
&~&~~~~~~~~~~~~\times \frac{1}{\Gamma_2 (p, \Omega)+r}.
\label{sp3}
\end{eqnarray}

Finally, we will also need the $1/N_f$ correction to the boson-fermion vertex (the Yukawa couping) in Eq.~(\ref{dsid4}). At zero
external momenta and frequencies, this vertex is renormalized by Fig.~\ref{fig:feyn2} to
\begin{eqnarray}
\Xi &=& \tau^x + \frac{1}{N_f} \int \frac{d^2 p}{4 \pi^2} \int \frac{d \Omega}{2 \pi}  
\Bigl[  \tau^x \left( - i \Omega + v_F p_x  \tau^z + v_\Delta p_y \tau^x \right)^{-1}  
\nonumber \\
&~&~~~~~~~~~~~\times \tau^x  \left( - i \Omega + v_F p_x  \tau^z + v_\Delta p_y \tau^x \right)^{-1} \tau^x \frac{1}{\Gamma_2 (p, \Omega) + r}  
\Bigr]
\label{sp4}
\end{eqnarray}

The RG equations will be obtained in the next section from Eqs.~(\ref{sp2}), (\ref{sp3}) and (\ref{sp4}).

\section{Renormalization group analysis}
\label{sec:rg}

We begin our discussion of the RG by describing the general structure which applies to {\em all\/} orders
in the $1/N_f$ expansion. The RG will be based upon a consideration of the renormalization of the local field
theory $S$ in Eq.~(\ref{local}). We are interested in the behavior of this field theory under the rescaling
transformation 
\begin{eqnarray}
k &=& k' e^{-\ell} \nonumber \\
\omega &=& \omega' e^{- \ell} .
\end{eqnarray}
Note that we have not introduced a dynamic critical exponent $z$ for the rescaling of the frequency.
This is because we will allow both velocities $v_F$ and $v_\Delta$ to flow, and this flow will effectively
account for any anomalous dynamic scaling. 
Under this spacetime rescaling, we also have a rescaling of the fermion and boson fields
\begin{eqnarray}
\Psi_{1,2} (k, \omega) &=& \Psi_{1,2}^\prime (k', \omega') \exp \left( \frac{1}{2} \int_0^\ell d \ell (4 - \eta_f) \right) \nonumber \\
\phi (k, \omega) &=& \phi^\prime (k', \omega')\exp \left( \frac{1}{2} \int_0^\ell d \ell (5 - \eta_b) \right).
\end{eqnarray}
Here we have allowed the anomalous dimensions $\eta_{b,f}$ to be scale-dependent, as that will be the case below.
To implement these field rescalings, we have to determine how the field scales are defined.
For the fermions, as is conventional, we set the scale of $\Psi_{1,2}$ so that the co-effecients
of the time derivative terms in $S_\Psi$ remain unity. For the nematic order parameter, $\phi$, there is no
``kinetic energy'' term in Eq.~(\ref{local}), and so we cannot use the conventional method. Instead, recall
that the ``Yukawa'' coupling $\lambda_0$ was absorbed into the overall scale of $\phi$ in Eq.~(\ref{local}).
Therefore it is natural to set the scale of $\phi$ so that the boson-fermion vertex in Eq.~(\ref{local}) remains
fixed at unity. This is also consonant with the fact that the boson ``kinetic energy'' comes entirely from the 
fermion loops.

The RG equations now follow from the low frequency forms of the fermion self energy $\Sigma_1$
and the vertex $\Xi$. These will have a dependence upon an ultraviolet cutoff, $\Lambda$. We will discuss
below an explicit method of applying this cutoff. For now, we note that for the RG we only need the logarithmic
derivative of the self energy and the vertex. Thus let us write
\begin{eqnarray}
\Lambda \frac{d}{d \Lambda} \Sigma_1 (k, \omega)  &=& C_1 (- i \omega) + C_2 v_F k_x \tau^z + C_3 v_\Delta k_y
\tau^x \nonumber \\
\Lambda \frac{d}{d \Lambda} \Xi  &=& C_4 \tau^x . \label{rg1}
\end{eqnarray}
On the right hand side, we assume (as in the usual field-theoretic RG) that the limit $\Lambda \rightarrow \infty$
can be safely taken. Then a simple dimensional analysis shows that the $C_{1-4}$ are all universal dimensionless
functions of the velocity ratio $v_\Delta/v_F$. In principle, the above method can be generalized to any needed order in the 
$1/N_f$ expansion by taking the appropriate logarithmic cutoff derivative of the self energy $\Sigma_1$ and the
vertex $\Xi$, as reviewed in Ref.~\onlinecite{brezin}.

With the results in Eq.~(\ref{rg1}), we can set the scaling dimensions of the fields. Considering the
renormalization of the $-i \omega$ term in $S_\Psi$, we obtain
\begin{equation}
\eta_f = -C_1.
\end{equation}
Imposing the unit value of the Yukawa coupling in Eq.~(\ref{local}) we obtain
\begin{equation}
\eta_b = 1 + 2 C_4 - 2 \eta_f = 1 + 2 C_4 + 2 C_1
\end{equation}
Note the large value of $\eta_b$ in the large $N_f$ theory. This is a consequence of the fact that the
$\phi$ kinetic energy is tied to the non-analytic fermion loop contribution. We can now also use the non-renormalization
of the $\phi^2$ term in Eq.~(\ref{local}), as noted in Eq.~(\ref{gamma}), to determine the flow equation for the 
``mass'' $r$. This defines the correlation length exponent $\nu$ by the RG equation
\begin{equation}
\frac{dr}{d \ell} = \frac{1}{\nu} r
\label{eqr}
\end{equation}
for small $r$, and we have the non-mean-field value
\begin{equation}
\nu = \frac{1}{2 - \eta_b}.
\end{equation}

The results in Eq.~(\ref{rg1}) also yield the flow equations for the velocities. 
By considering the renormalization of the $k$-dependent terms in $S_\Psi$ we determine that
the velocity $v_F$ flows according 
\begin{equation}
\frac{d v_F} {d \ell} = (-\eta_f - C_2) v_F = (C_1 - C_2) v_F , \label{vfl}
\end{equation}
while the velocity $v_\Delta$ obeys
\begin{equation}
\frac{d v_\Delta}{d \ell} = (- \eta_f - C_3) v_\Delta = (C_1 - C_3) v_\Delta . \label{vdl}
\end{equation}
Clearly, the ratios of the velocity obeys
\begin{equation}
\frac{d (v_\Delta/v_F)}{d \ell} = (C_2 - C_3) (v_\Delta/v_F). \label{vdfl}
\end{equation}

All that remains now is to determine the $C_{1-4}$ as a function of $v_\Delta /v_F$. 
These functions also depend upon the ``mass'' $r$, but we will henceforth restrict ourselves to the 
vicinity of the critical point by setting $r=0$. 
We reiterate that, in principle, our method allows us to obtain these functions to any needed order in the $1/N_f$ expansion.
However, we will only obtain explicit results below to first order $1/N_f$.

The computation of $C_{1-4}$ requires the imposition of a ultraviolet momentum cutoff. We implement this \cite{vojta}
by multiplying both the Fermi and Bose propagators by a smooth cutoff function $\mathcal{K} (k^2 /\Lambda^2)$.
Here $\mathcal{K}(y)$ is an arbitrary function with $\mathcal{K}(0)=1$ and which falls off rapidly with $y$ at $y \sim 1$
{\em e.g.\/} $\mathcal{K} (y) = e^{-y}$. We will show below that the RG equations are independent of the particular choice
of $\mathcal{K}(y)$.

Let us now discuss the evaluation of $C_{1-3}$. Schematically, we can write for the fermion self energy
\begin{equation}
\Sigma_1 (K) = \int \frac{d^3 P}{8 \pi^3} F(P+K) G (P) \mathcal{K} \left(\frac{(p+k)^2}{\Lambda^2} \right)
\mathcal{K} \left(\frac{p^2}{\Lambda^2} \right) \label{rg0}
\end{equation}
where $K \equiv (k, \omega)$ and $P \equiv (p, \Omega)$ are 3-momenta, and $F$ and $G$ are
functions which can be obtained by comparing this expression to Eq.~(\ref{sp3}). Also notice that
(at $r=0$) $F$ and $G$ are both homogenous functions of 3-momenta of degree -1.
Expanding to first order in $K_\mu$ ($\mu$ is a spacetime index) we have
\begin{equation}
\Sigma_1 (K) \approx K_\mu \int \frac{d^3 P}{8 \pi^3}\left[ \frac{\partial F (P) }{\partial P_\mu} G (P) \mathcal{K}^2 \left(\frac{p^2}{\Lambda^2} \right)  + F(P) G (P) \mathcal{K} \left(\frac{p^2}{\Lambda^2} \right) \frac{2 p_\mu}{\Lambda^2} \mathcal{K}' \left(\frac{p^2}{\Lambda^2} \right) \right]
\end{equation}
where $p_\mu$ is the three vector $(0,p_x,p_y)$ (while $P_\mu = (\Omega, p_x, p_y)$).
Now taking the $\Lambda$ derivative
\begin{eqnarray}
\Lambda \frac{d}{d\Lambda} \Sigma_1 (K) &\approx& K_\mu \int \frac{d^3 P}{8 \pi^3}\left[\left\{ - \frac{4p^2}{\Lambda^2} \frac{\partial F (P) }{\partial P_\mu} - 4 F(P) \frac{p_\mu}{\Lambda^2} \right\} G (P)  \mathcal{K} \left(\frac{p^2}{\Lambda^2} \right) \mathcal{K}' \left(\frac{p^2}{\Lambda^2} \right) \right. \nonumber \\
 &~& \left.~~- \frac{4 p^2 p_\mu}{\Lambda^4} F(P) G (P) \left\{
 \mathcal{K} \left(\frac{p^2}{\Lambda^2} \right)  \mathcal{K}'' \left(\frac{p^2}{\Lambda^2} \right) + \mathcal{K}^{\prime 2} \left(\frac{p^2}{\Lambda^2} \right)\right\} \right]
\end{eqnarray}
Now we convert to cylindrical co-ordinates in spacetime by writing $P_\mu = y \Lambda  (v_F x, \cos \theta , \sin \theta)$
and integrate over $y$, $x$, and $\theta$. Also, let us define $\hat{P}_\mu = (v_F x, \cos \theta, \sin \theta)$ and
$\hat{p}_\mu = (0, \cos \theta, \sin \theta)$.
We use the homogeneity properties of the functions $F$ and $G$ 
to obtain
\begin{eqnarray}
\Lambda \frac{d}{d\Lambda} \Sigma_1 (K) &\approx& \frac{v_F K_\mu}{8 \pi^3} \int_{-\infty}^\infty dx \int_0^{2 \pi} d \theta \left[ \left\{- 4
 \frac{\partial F (\hat{P}) }{\partial P_\mu} - 4 \hat{p}_\mu F(\hat{P}) \right\} G (\hat{P}) \int_0^\infty y dy \mathcal{K}(y^2) \mathcal{K}' (y^2)   \right. \nonumber \\
 &~& \left.~~- 4 \hat{p}_\mu  F(\hat{P}) G (\hat{P}) 
 \int_0^\infty y^3 dy 
 \left\{
 \mathcal{K} \left(y^2 \right)  \mathcal{K}'' \left(y^2 \right) + \mathcal{K}^{\prime 2} \left(y^2 \right)\right\} \right]
\end{eqnarray}
Now the integrals over $y$ can be evaluated exactly by integration by parts, and the final result is
{\em independent\/} of the precise form of $\mathcal{K}(y)$:
\begin{equation}
\Lambda \frac{d}{d\Lambda} \Sigma_1 (K) = \frac{v_F K_\mu}{8 \pi^3} \int_{-\infty}^\infty dx \int_0^{2 \pi} d \theta 
 \frac{\partial F (\hat{P}) }{\partial P_\mu} G (\hat{P}) \label{rg2}
 \end{equation}
 
In a similar manner, we can write the result for the vertex $\Xi$ in Eq.~(\ref{sp4}) in the schematic form
\begin{equation}
\Xi = \tau^x + \int \frac{d^3 P}{8 \pi^3} H (P) \mathcal{K}^3 \left(\frac{p^2}{\Lambda^2} \right), \label{rg7}
\end{equation}
where $H(P)$ is a homogenous function of $P$ of degree -3.
Proceeding as above we obtain the analog of Eq.~(\ref{rg2})
\begin{equation}
\Lambda \frac{d}{d\Lambda} \Xi  = \frac{v_F}{8 \pi^3} \int_{-\infty}^\infty dx \int_0^{2 \pi} d \theta 
 H (\hat{P}). \label{rg3}
 \end{equation}

\begin{figure}[h]
\includegraphics[width=4in]{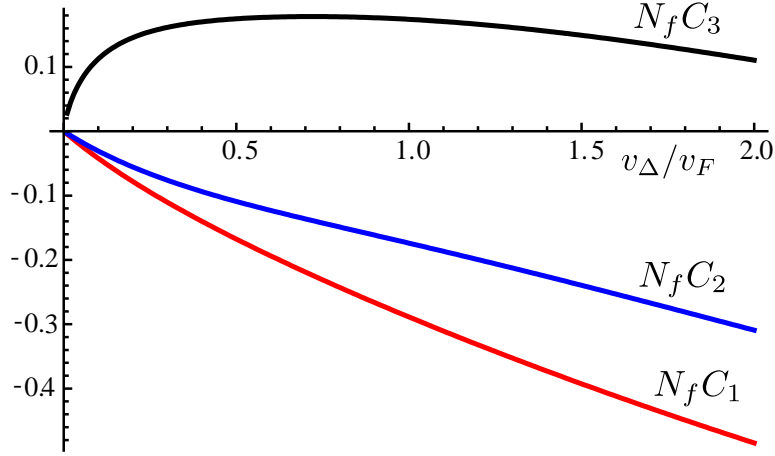}
\caption{Plots of the functions $C_{1-3}$ in Eq.~(\ref{rg5}).} \label{fig:cplot}
\end{figure}

We can now use the above expressions to obtain explicit results for the constants $C_{1-4}$ at order
$1/N_f$. For $C_{1-3}$ we combine Eqs.~(\ref{sp3}), (\ref{rg1}), (\ref{rg0}), and (\ref{rg2}) to obtain
\begin{eqnarray}
C_1 &=&  \frac{2(v_\Delta/v_F)}{\pi^3 N_f} \int_{-\infty}^\infty dx \int_0^{2 \pi} d \theta 
 \frac{(x^2 -  \cos^2 \theta - (v_\Delta/v_F)^2 \sin^2 \theta)}{(x^2 + \cos^2 \theta + (v_\Delta/v_F)^2 \sin^2 \theta)^2} \mathcal{G} (x, \theta)
\nonumber \\
C_2 &=&  \frac{2(v_\Delta/v_F)}{\pi^3 N_f} \int_{-\infty}^\infty dx \int_0^{2 \pi} d \theta 
 \frac{(-x^2 +  \cos^2 \theta - (v_\Delta/v_F)^2 \sin^2 \theta)}{(x^2 + \cos^2 \theta + (v_\Delta/v_F)^2 \sin^2 \theta)^2} \mathcal{G} (x, \theta)
\nonumber \\
C_3 &=&  \frac{2(v_\Delta/v_F)}{\pi^3 N_f} \int_{-\infty}^\infty dx \int_0^{2 \pi} d \theta 
 \frac{(x^2 + \cos^2 \theta - (v_\Delta/v_F)^2 \sin^2 \theta)}{(x^2 + \cos^2 \theta + (v_\Delta/v_F)^2 \sin^2 \theta)^2} \mathcal{G} (x, \theta)
\label{rg5}
\end{eqnarray}
where
\begin{equation}
\mathcal{G}^{-1} (x, \theta) = \frac{x^2 + \sin^2 \theta}{\sqrt{
x^2 + (v_\Delta/v_F)^2 \cos^2 \theta + \sin^2 \theta}} +   \frac{x^2 + \cos^2 \theta}{\sqrt{
x^2 + \cos^2 \theta + (v_\Delta/v_F)^2 \sin^2 \theta}}
\end{equation}
is the $\phi$ propagator inverse.
All the integrals in Eqs.~(\ref{rg5}) are well-defined and convergent for all non-zero $v_\Delta/v_F$, and have to be 
evaluated numerically. It is also evident that the results are functions only of $v_\Delta/v_F$. The results of a numerical evaluation
are shown in Fig.~\ref{fig:cplot}.

Similarly, we can combine Eqs.~(\ref{sp4}), (\ref{rg1}), (\ref{rg7}), and (\ref{rg3}) to deduce that
\begin{equation}
C_4 = - C_3,
\end{equation}
at leading order in the $1/N_f$ expansion.

\begin{figure}[h]
\includegraphics[width=4in]{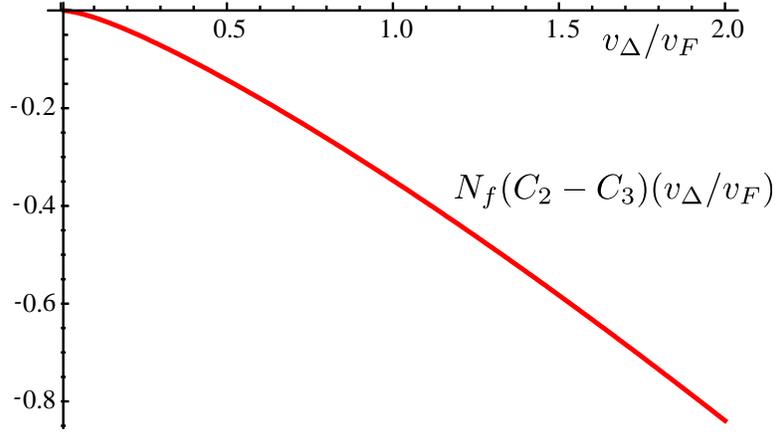}
\caption{Right-hand-side of the RG flow equation Eq.~(\ref{vdfl}) for $v_\Delta/v_F$.} \label{fig:betaplot}
\end{figure}

With the values in Eq.~(\ref{rg5}), we are now in a position to numerically solve the RG flow for the velocity ratio $v_\Delta/v_F$
in Eq.~(\ref{vdfl}). We plot the right-hand-side of this flow equation in Fig.~\ref{fig:betaplot}.
Note that the only zero of the flow equation is at $v_\Delta/v_F = 0$, and that this fixed point is attractive in the infrared. So for all starting values of the parameter $v_\Delta/v_F$, the flow is towards $v_\Delta/v_F
\rightarrow 0$ for large $\ell$. At large enough scales, it is always possible to approximate the constants $C_{1-3}$
in Eqs.~(\ref{rg5}) by their limiting values for small $v_\Delta /v_F$. Evaluating the integrals in Eq.~(\ref{rg5}) in 
this limit, we obtain
\begin{eqnarray}
C_1 &=& -0.4627 \frac{(v_\Delta/v_F)}{N_f} + \mathcal{O} \left((v_\Delta/v_F)^3 \right) \nonumber \\
C_2 &=& -0.3479 \frac{(v_\Delta/v_F)}{N_f} + \mathcal{O} \left((v_\Delta/v_F)^3 \right) \nonumber \\
C_3 &=& \left( \frac{8}{\pi^2} \ln (v_F/v_\Delta) - 0.9601 \right) \frac{(v_\Delta/v_F)}{N_f} + \mathcal{O} \left((v_\Delta/v_F)^3 \right) .\label{cres}
\end{eqnarray}
The logarithmic dependence on $v_\Delta/v_F$ arises from a singularity in the integrand at $x=0$ and $\theta = \pm \pi/2$: this
corresponds to an integral across the underlying Fermi surface over $\omega$ and $k_x$ where the inverse
fermion propagator $\sim -i \omega + v_F k_x \tau^z$. 
Inserting these values into the flow equations (\ref{vfl}), (\ref{vdl}), and (\ref{vdfl}) we obtain an explicit form of
the RG flow for small $v_\Delta/v_F$:
\begin{eqnarray}
\frac{d v_\Delta}{d \ell} &=& -  \left( \frac{8}{\pi^2}  \ln \left({v_F}/{v_\Delta} \right) -0.4974 \right) \frac{v_\Delta^2}{N_f v_F} \label{rgf1} \\
\frac{d v_F}{d \ell} &=&    -0.1148 \frac{v_\Delta}{N_f}  \label{rgf2} \\
\frac{d (v_\Delta/v_F) }{d \ell} &=&    -  \left( \frac{8}{\pi^2}  \ln \left({v_F}/{v_\Delta} \right) -0.6122 \right) \frac{(v_\Delta/v_F)^2}{N_f} .
\label{rgf3}
\end{eqnarray}
The right-hand-side of Eq.~(\ref{rgf3}) has a zero for $v_\Delta/v_F \approx 2$; this is spurious as this equation
is valid only for $v_\Delta/v_F \ll 1$. The full expression for the right-hand-side, valid for arbitrary $v_\Delta/v_F$, is plotted in Fig.~\ref{fig:betaplot}, and this
makes it clear that the only zero of the beta function is at $v_\Delta/v_F = 0$.

\begin{figure}[h]
\includegraphics[width=3.1in]{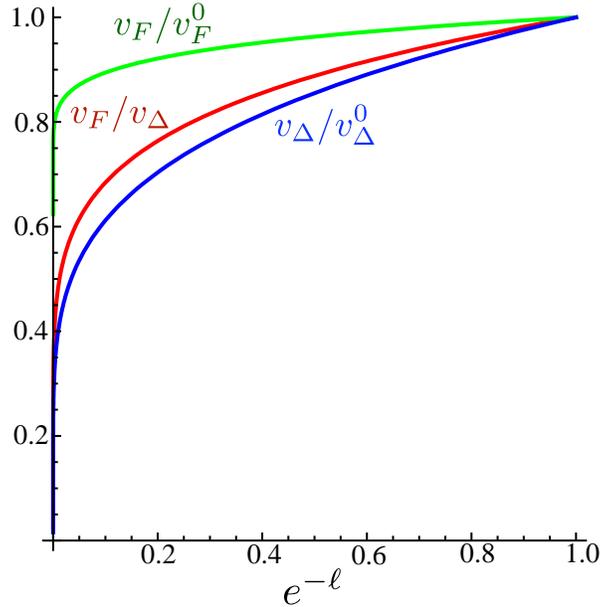}
\caption{Integration of Eqs.~(\ref{vfl}), (\ref{vdl}) and (\ref{vdfl}) from $\ll = 0$ to $\ell$ for $N_f=2$,
starting from the velocities $v_F^0$ and $v_\Delta^0$. Results above are for $v_\Delta^0/v_F^0 = 1$.} \label{fig:vplot1}
\end{figure}

\begin{figure}[h]
\includegraphics[width=3.1in]{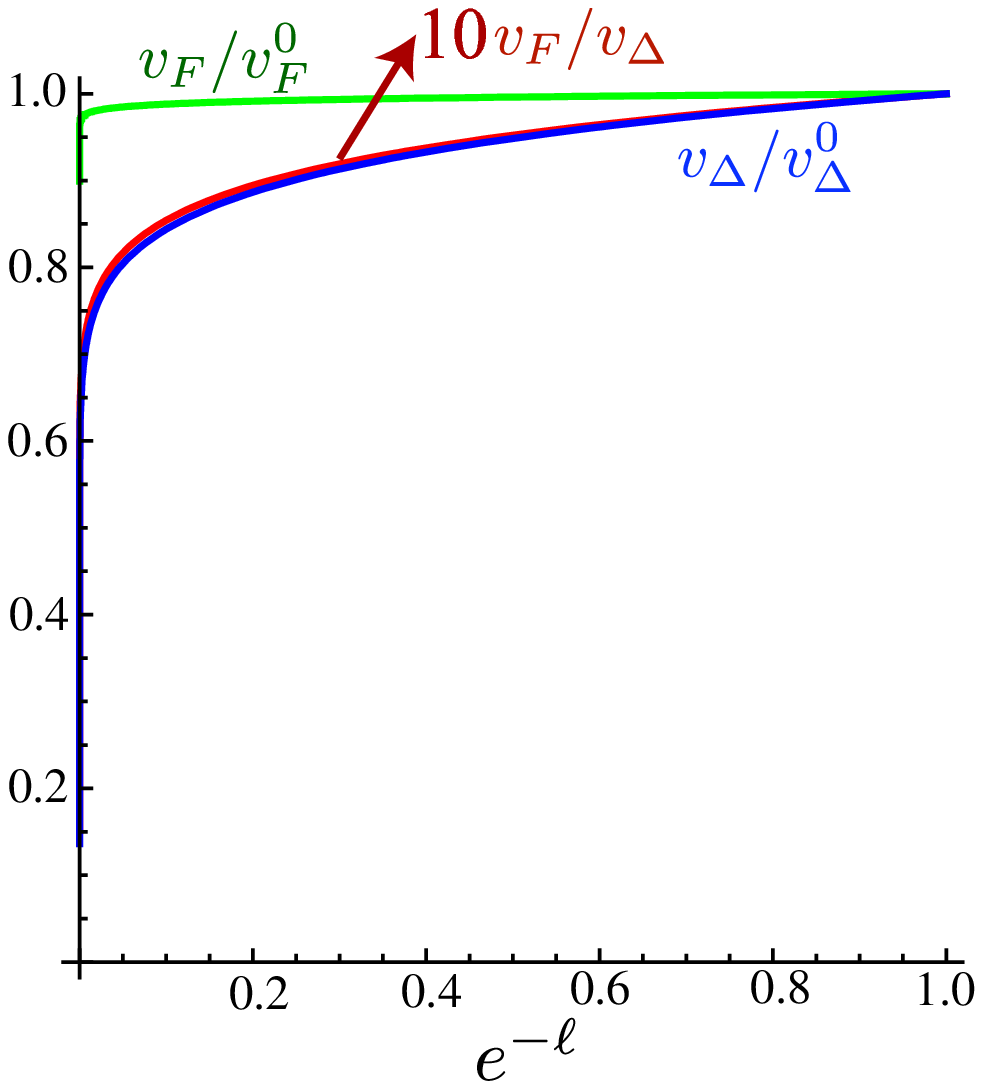}
\caption{As in Fig.~\ref{fig:vplot1} but for $v_\Delta^0 /v_F^0 = 0.1$.} \label{fig:vplot2}
\end{figure}

\begin{figure}[h]
\includegraphics[width=3.1in]{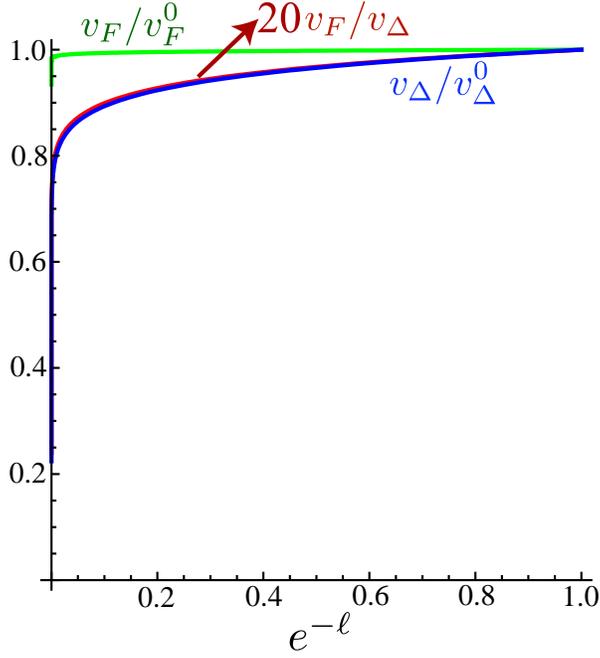}
\caption{As in Fig.~\ref{fig:vplot1} but for $v_\Delta^0 /v_F^0 = 0.05$.} \label{fig:vplot3}
\end{figure}

The results of a numerical integration of Eqs.~(\ref{vfl}), (\ref{vdl}) and (\ref{vdfl}) are shown in Figs.~\ref{fig:vplot1},
\ref{fig:vplot2}, and \ref{fig:vplot3}.
We start the RG integration at $\ell = 0$ corresponding to a temperature $T=T_0$ at the nematic critical
point,
at which point the velocities have the values $v_F^0$ and $v_\Delta^0$. Integrating to the 
scale $\ell$ yields the velocities $v_F (\ell)$ and $v_\Delta (\ell)$ at a temperature $T = T_0
e^{-\ell}$. The results depend upon the initial value of the velocity ratio, $v_\Delta^0 /v_F^0$,
and are shown in the figures for $v_\Delta^0 /v_F^0 = 1, 0.1, 0.05$. These results can also be converted
into dependence on the deviation from the nematic critical point, $r$, by integrating Eq.~(\ref{eqr}).

In the asymptotic low temperature regime ($\ell \rightarrow \infty$), we  have $v_\Delta/v_F \rightarrow 0$,
and so we can integrate the asymptotic expression in Eq.~(\ref{rgf3}). This yields
\begin{equation}
\frac{v_\Delta}{v_F} = \frac{\pi^2 N_f}{8} \frac{1}{\ell} \frac{1}{ \ln[0.3809 \, \ell/N_f]} \label{vdvfres}
\end{equation}
The $T$ dependence follows by using $\ell = \ln (T_0/T)$. Using the result for $v_\Delta/v_F$ in Eq.~(\ref{vdvfres})
and integrating Eq.~(\ref{rgf2}), we find that $v_F$ only has a finite renormalization from its bare value as $\ell \rightarrow \infty$.
So the decrease in $v_\Delta/v_F$ as $\ell \rightarrow \infty$ ({\em i.e.\/} as $T \rightarrow 0$) comes primarily
from the decrease in $v_\Delta$. This is also clear from the plots in Figs.~\ref{fig:vplot1}-\ref{fig:vplot3}.

We can also apply these results to obtain the $T=0$ evolution of the velocities as a function of the tuning
parameter, $r$, across the nematic ordering transition. In this case we use Eq.~(\ref{eqr}), along with
the fact that $\nu \rightarrow 1$ as $v_\Delta /v_F \rightarrow 0$. Then we deduce that $\ell = \ln (1/|r|)$ to leading
logarithm accuracy,
and so obtain the $r$ dependence of the velocities from Figs.~\ref{fig:vplot1}-\ref{fig:vplot3}. 
When both $T$ and $r$ are non-zero, we can use $\ell = \ln (\max(|r|,T))$ to leading logarithmic accuracy.
Note that this predicts a minimum in $v_\Delta/v_F$ as $r$ is tuned across the nematic ordering transition, with
the minimum value of order $1/\ln(1/T)$.

\section{Theory for small $v_\Delta /v_F$}
\label{sec:ho}

The RG analysis in Section~\ref{sec:rg} has shown that the $\beta$ functions at order $1/N_f$ 
are such that the flow is towards the strong anisotropy limit, $v_\Delta /v_F \rightarrow 0$ at large scales.
Here we will show that this conclusion holds at all orders in the $1/N_f$ expansion. Indeed, for small $v_\Delta/v_F$
we will show that $v_\Delta/v_F$ can itself be as the control parameter for the computation, and $N_f$ is not
required to be large. In other words, the RG flow equations in Eqs.~(\ref{rgf1}-\ref{rgf3}) are asymptotically {\em exact\/}, even
for the physical value of $N_f = 2$. All corrections to Eq.~(\ref{rgf1}-\ref{rgf3}) are higher order in $v_\Delta /v_F$.

Our conclusions follow from an examination of the structure of the action $S_\phi$ for the nematic order
in Eq.~(\ref{sp1}) in the limit of $v_\Delta/v_F \rightarrow 0$. As we will shortly verify, the natural scale
for the fluctuations of $\phi$ are at frequency and momenta  
with $\omega \sim v_F k_x \sim v_F k_y$. Under these conditions,
we can just take the $v_\Delta \rightarrow 0$ limit of the expressions in Eqs.~(\ref{pi2r}), (\ref{p4a}), 
and (\ref{p4b}). This limit exists, and to quadratic order in $\phi$ we have the following effective action as $v_\Delta
\rightarrow 0$ (at criticality, $r=0$):
\begin{equation}
S_\phi = \frac{N_f}{16 v_F v_\Delta} \int \frac{d^2 k}{4 \pi^2} \int \frac{ d \omega}{2 \pi}
|\phi (k, \omega)|^2 \left[ \sqrt{\omega^2 + v_F^2 k_x^2} + \sqrt{\omega^2 + v_F^2 k_y^2} \right] + \ldots
\label{ex1}
\end{equation}
Rescaling momenta $k \rightarrow k/v_F$, we notice that the prefactor in Eq.~(\ref{ex1}) is proportional to the 
dimensionless number
$N_f v_F /v_\Delta$. Thus each propagator of $\phi$ will appear with the small prefactor of $v_\Delta /(N_f v_F)$.
Further, as claimed above, each propogator has the typical scales $\omega \sim v_F k_x \sim v_F k_y$.

These considerations are easily extended to terms to all orders in $\phi$ in Eq.(\ref{sp1}).  The full expression 
for $S_\phi$ involves the sum of the logarithms of the determinants of the Dirac operators of $\Psi_{1,2}$ fermions
in a background $\phi$ field. The expansion of these determinants in powers of $\phi$ involves only terms
with a single fermion loop. For each $\Psi_1$ loop we rescale the {\em fermion\/} 
momentum $p_x \rightarrow p_x /v_F$
and $p_y \rightarrow p_y /v_\Delta$. Similarly for each  $\Psi_2$ loop we rescale the fermion 
momentum $p_x \rightarrow p_x /v_\Delta$
and $p_y \rightarrow p_y /v_F$. In both cases, the integral over the fermion loop momentum yields
a prefactor of $1/(v_F v_\Delta)$. For the $\phi$ vertices emerging from the $\Psi_1$ loops, we have 
momenta $\omega \sim v_F k_x \sim v_F k_y$, as noted above; when this external momenta enters
the fermion loop, we have $k_x \sim p_x$, the typical $\Psi_1$ $x$-momentum. 
However, $k_y \ll p_y \sim 1/v_\Delta$, the typical $\Psi_1$ $y$-momentum. Consequently, we may safely
set $k_y = 0$ within the fermion loop, and there is no further dependence upon $v_\Delta$ in the loop integral.
Similarly, in the $\Psi_2$ loops, we may set $k_x =0$ within the fermion loop. To summarize, the typical
$\phi$ momenta are $k_x \sim 1/v_F$, $k_y \sim 1/v_F$, the typical $\Psi_1$ momenta are $p_x \sim 1/v_F$, 
$p_y \sim 1/v_\Delta$, and typical $\Psi_2$ momenta are $p_x \sim 1/v_\Delta$, 
$p_y \sim 1/v_F$. Further, the resulting effective action for $\phi$ has the form
\begin{equation}
S_\phi = \frac{N_f}{v_F v_\Delta} \mathcal{S} \bigl[ \phi (k, \omega); v_F k_x, v_F k_y, \omega \bigr] + \mbox{
terms higher order in $v_\Delta/v_F$.}
\label{ex2}
\end{equation}
Here $\mathcal{S}$ is a functional of $\phi$ which is independent of $v_\Delta$, whose expansion in $\phi$
has co-efficients which depend only upon $v_F k_x$, $v_F k_y$, and $\omega$. It can be verified
that Eq.~(\ref{ex1}) and the quartic terms in Eqs.~(\ref{p4a}) and (\ref{p4b}) are indeed of the form 
in Eq.~(\ref{ex2}).
 
From Eq.~(\ref{ex2}), it is evident (after rescaling momenta $k \rightarrow k/v_F$ and momentum space
fields $\phi \rightarrow v_F^2 \phi$) that the natural expansion
parameter controlling $\phi$ fluctuations is $v_\Delta/(N v_F)$. This is the main result of this section.

It remains to understand the $\ln (v_F/v_\Delta)$ factor in Eq.~(\ref{rgf3}).  If we compute observable $\phi$
correlations under the reduced effective action $\mathcal{S}$, the resulting expansion in powers of $v_\Delta/(N v_F)$
leads to Feynman integrals which are not necessarily infrared or ultraviolet finite. 
Because the action $\mathcal{S}$ is free
of any dimensionful coupling constant, the Feynman graph divergences can at most be powers of logarithms.
These divergences are cutoff by using the full fermion propagators, including the terms proportional to $v_\Delta$
times the momenta of the boson propagators. This leads to factors of $\ln (v_F /v_\Delta)$, as was the case
in obtaining Eq.~(\ref{rgf3}) from Eq.~(\ref{rg5}).

\section{Conclusions}
\label{sec:conc}

This paper has described the RG properties of the field theory \cite{vojta} in Eq.~(\ref{local}) for nematic ordering. It was previously noted \cite{eakim} that the effective potential for the nematic order, $\phi$,
remained unrenormalized upon integrating out the nodal fermions $\Psi_{1,2}$. This happens because
the $\phi$ couples to a conserved fermion current, and a spacetime-independent $\phi$ can be 
`gauged away' applying a gauge transformation to the fermions. One consequence of this non-renormalization
is that the susceptibility exponent, $\gamma$, takes the simple value in Eq.~(\ref{gamma}). 
However, non-trivial renormalizations of the field scale of both $\phi$ and $\Psi_{1,2}$ are still possible,
along with renormalizations of the velocities, and we have shown here that this leads to an interesting
RG flow structure, with non-mean-field exponents.

It is interesting to note here the similarities to the RG flow of supersymmetric field theories \cite{strassler}.
There the effective potential also remains unrenormalized (albeit for very different reasons), but the 
`wavefunction renormalizations' lead to many non-trivial RG fixed points.

Our main result was that the transition is described by a fixed point in which the fermion velocities
at the nodal points have a ratio which approaches a fixed point with $v_\Delta /v_F \rightarrow 0$
logarithmically slowly. However, it is not valid
to set the pairing-induced velocity $v_\Delta = 0$ in the computation, and so deal with a metallic
Fermi surface. For the case where the fermion dispersion is $\sim (v_F^2 k_x^2 + v_\Delta^2 k_y^2)^{1/2}$,
the typical fermion momenta contributing to the critical theory scale as $k_x \sim 1/v_F$ and $k_y \sim
1/v_\Delta$, and so the full form of the Bogoliubov quasi-particle dispersion is important.
The flow of the velocities is described by Eqs.~(\ref{rgf1}-\ref{rgf3}) for $v_\Delta/v_F \ll 1$, and these equations are believed
to be asymptotically exact.

Unlike the fermions, the fluctuations of the nematic order, $\phi$, are isotropic in momenta. These are described asymptotically
exactly by Eq.~(\ref{ex1}). Note that the propagator is very different from free field, and has a large anomalous
dimension, as was found in the $N_f= \infty$ theory \cite{eakim}. 
Experimental detection of this unusual spectrum in {\em e.g.\/} inelastic X-ray scattering would be 
most interesting.

\acknowledgments We thank M.~P.~A.~Fisher, E.~Fradkin, E.-A.~Kim, S.~Kivelson, and M.~Lawler for useful discussions. This research was supported by the NSF under
grant DMR-0537077.

\end{document}